\documentclass[a4paper,aps,pre,preprint,nofootinbib,showpacs,12pt]{revtex4-1}

\usepackage{graphicx}
\usepackage{amsfonts}
\usepackage{amsmath}
\usepackage{amsbsy}
\usepackage{longtable}
\usepackage{bm}
\usepackage{hyperref}
\usepackage{float}
\usepackage{epstopdf}
\allowdisplaybreaks

\begin{document}
\title{Shear viscosity of an ultrarelativistic Boltzmann gas with isotropic inelastic scattering
processes}

\author{A. El$^1$\footnote{el@th.physik.uni-frankfurt.de},
F. Lauciello$^1$,
C. Wesp$^1$,
I. Bouras$^1$\footnote{bouras@th.physik.uni-frankfurt.de},
Z. Xu$^{1,2,3}$\footnote{xu@th.physik.uni-frankfurt.de}, 
C. Greiner$^{1}$\footnote{Carsten.Greiner@th.physik.uni-frankfurt.de}}

\affiliation{$^1$ Institut f\"{u}r Theoretische Physik, 
Goethe-Universit\"{a}t Frankfurt,
Max-von-Laue Str. 1, D-60438, Frankfurt am Main, Germany}
\affiliation{$^2$ Department of Physics, Tsinghua University, Beijing 100084, China}
\affiliation{$^3$ Collaborative Innovation Center of Quantum Matter, Beijing, 100084 China}

\begin{abstract}

We derive an analytic expression for the shear viscosity of an ultra-relativistic gas in presence of
both elastic $2\to 2$ and inelastic $2\leftrightarrow 3$ processes with isotropic differential
cross sections. The derivation is based on the entropy principle and Grad's approximation for the
off-equilibrium distribution function. The obtained formula relates the shear viscosity coefficient
$\eta$ to the total cross sections $\sigma_{22}$ and $\sigma_{23}$ of the elastic resp. inelastic
processes. The values of shear viscosity extracted using the Green-Kubo formula from kinetic
transport calculations are shown to be in excellent agreement with the analytic results which
demonstrates the validity of the derived formula. 

\end{abstract}


\maketitle

\section{Introduction}

Dynamics of the hot and dense state of nuclear matter, produced in heavy-ion collisions at modern
accelerator facilities like RHIC at Brookhaven National Laboratory or LHC at CERN, can be
described, with exception of the very early times governed by off-shell dynamics,
by means of kinetic transport theory or relativistic hydrodynamics. In fact, these two approaches
have been widely used for research of the properties of the so-called Quark-Gluon Plasma (QGP),
early-time dynamics of which determines the properties of hadronic and electromagnetic
secondaries measured experimentally \cite{hydros,Xu:2004mz,Xu:2007jv}. Whereas the kinetic transport
theory is based on a
microscopic paradigm and describes evolution of the phase-space distribution function by solving
the corresponding transport equations, hydrodynamics is a macroscopic theory describing space and
time evolution of macroscopic fields like the energy-momentum tensor and particle number
four-vector. For the hydrodynamic theory the transport coefficients are an extrinsic input and must
be calculated from the underlying field theory, which for QGP is the Quantum Chromodynamics (QCD).
Calculations of the transport coefficients are based on the formal correspondence
between the kinetic theory and the hydrodynamics, which can be obtained by integration of the
microscopic equations over the phase space. For example, such correspondence is discussed in
\cite{IS,Muronga:2003ta,Baier:2006gy,Betz:2009zz,Denicol:2010xn}. Moreover, kinetic transport
calculations have often been used as a benchmark for hydrodynamic models, as for instance in Refs.
\cite{Huovinen:2008te,El:2009vj,Bouras:2010hm}, which again requires a consistent connection between
the cross section on the one hand and transport coefficients on the other.

For a strongly interacting QGP, the shear viscosity coefficient has been
calculated in the perturbative regime as function of the coupling parameter $\alpha_s$ for
instance in Refs. \cite{Arnold:2000dr}, \cite{El:2008yy,Wesp:2011yy,Reining:2011xn}. Such
calculations account for the radiative processes
occurring in the QGP, which have an important contribution to the isotropization of the medium. The
role of inelastic processes was understood by means of the transport cross section or transport
rate \cite{Xu:2007aa}, which are analogous to the total cross section and collision rate but are
additionally weighted by the angle of outgoing particles. In contrast, calculations reported in
Refs. \cite{Chen:2010xk} do not see a strong contributions of the inelastic $2\to 3$ processes. In
Ref. \cite{Fuini:2010xz} the shear viscosity of a QGP was calculated employing two-particle
scatterings, i.e. without taking into account the radiative processes. 

In this work we derive an expression which relates the shear viscosity coefficient of an
ultrarelativistic gas to the total elastic and inelastic cross sections in the medium.
For our derivation we consider isotropic, i.e. momentum-independent, matrix elements for the
collisional processes, thus bypassing the conceptual difficulties connected with the
momentum-dependent matrix elements in a (perturbative) QCD medium. To our best knowledge,
analytic derivation of the shear viscosity for $2\leftrightarrow 3$ inelastic  processes with
isotropic differential cross sections is completely new.

This paper has the following structure. In Section \ref{sec:derivation} we start with the derivation
of an expression for the shear viscosity coefficient, for which we use the entropy production in a
system off equilibrium. In order to obtain a closed analytic expression for the shear viscosity
coefficient in terms of the involved total cross sections, kinematic integrals will be evaluated.
This is discussed in Section \ref{sec:evaluation}, where the final expression is also presented. In
Section \ref{sec:comparison} the results of the analytic formula are compared with Green-Kubo based
kinetic transport calculations of the shear viscosity coefficient. Finally, Section
\ref{sec:conclusion} contains summary and conclusions. 

\section{Derivation of the expression for the shear viscosity coefficient}\label{sec:derivation}

The derivation we follow is built upon the entropy principle, i.e. the second law of
thermodynamics is used. In presence of dissipative fields, of which we will consider only the
shear stress tensor throughout this work, the entropy production $\partial_\mu s^\mu$ must have a
non-negative form, and hence can be written as discussed e.g. in Ref. \cite{El:2010mt}: 
\begin{equation}
 \partial_\mu s^\mu = \frac{\tau_{\mu\nu}\tau^{\mu\nu}}{2\eta T} + J \ln \lambda \ge 0 \,.
\label{eq:dmusmu1}
\end{equation}
In the latter expression $\tau^{\mu\nu}$ denotes the shear tensor and $\eta$ the shear viscosity
coefficient; $T$ is the temperature and $\lambda$ the fugacity, which are both related to the
chemical potential, $\ln\lambda\equiv \frac{\mu}{T}$. $J$ denotes a source for particle production
and annihilation and vanishes in the situation where the net particle number does not change.

Note that
$\eta$ in Eq. (\ref{eq:dmusmu1}) is identified with the shear viscosity since one can show that in
the lowest-order, i.e. Navier-Stokes, dissipative hydrodynamic theory, where $\tau^{\mu\nu} = 2\eta
\nabla^{\langle \mu} u^{\rangle\nu}$, the entropy production indeed takes the form
(\ref{eq:dmusmu1}). Since higher-order theories (such as Israel-Stewart \cite{IS}) are supposed to
converge to the Navier-Stokes form, the entropy production is always written in the form 
(\ref{eq:dmusmu1}) and a consistent dynamic equation for  $\tau^{\mu\nu}$ can be found, like e.g. in
Ref. \cite{Muronga:2003ta}.

On the other hand, the entropy production can be obtained from the total divergence of the entropy
current defined in the kinetic theory as
\begin{equation}
s^\mu = - \int p^\mu f(x,p) \left[ \ln f(x,p) - 1 \right] d\Gamma \,,\label{eq:smu_kin} 
\end{equation}
where $f(x,p)$ denotes the phase-space distribution function and $d\Gamma\equiv d^3p/(2\pi)^3/E$.
For systems away from kinetic equilibrium the distribution can be written introducing a small
deviation $\phi(x,p)$ from the equilibrium form $f_0(x,p)$:
\begin{equation}
 f(x,p) = f_0(x,p)\cdot\left( 1 + \phi(x,p) \right) \,.\label{eq:grad}
\end{equation}
Some constrains must be applied to Eq. (\ref{eq:grad}) in order to preserve consistency. For
instance, if moments of $f(x,p)$ from (\ref{eq:grad}) are calculated, the particle number
four-vector $N^\mu$ and the energy-momentum tensor $T^{\mu\nu}$ must be recovered. This means that
the deviation $\phi(x,p)$ must depend on the deviations $\delta T^{\mu\nu}$ and $\delta N^\mu$ of
the energy-momentum tensor resp. particle number four-vector from their ideal forms. The explicit
functional dependence is given by the following expression, as obtained in scope of Grad's method of
moments \cite{Grad49,DeGroot,El:2008yy,Muronga:2006zx}:
\begin{equation}
 \phi(x,p) = c(e,p,T) \cdot \tau_{\mu\nu}p^\mu p^\nu \,.
\label{eq:phi}
\end{equation}
Note that the momentum-dependence of $\phi$ can be extended beyond the second power, as
demonstrated in Ref. \cite{Denicol:2012cn}. However, our derivation is restricted to the  original
approach of Grad, for which the series must be truncated at second power of momentum. In the kinetic
theory, dynamics of $f(x,p)$ is governed by a transport equation, such as the Boltzmann Equation 
\begin{equation}
p^\mu \partial_\mu f(x,p) = \mathcal{C}\left[ f(x,p) \right] \,,\label{eq:BE}
\end{equation}
with the functional $\mathcal{C}\left[ f(x,p) \right]$ on the right hand side denoting the collision
term. Combining Eqs. (\ref{eq:smu_kin}), (\ref{eq:grad}), (\ref{eq:BE}) and (\ref{eq:dmusmu1}) one
obtains the following expression for the shear viscosity coefficient \cite{El:2008yy,Muronga:2006zx}
\begin{equation}
 \eta = \frac{\tau_{\mu\nu}\tau^{\mu\nu}}{2 c(e,p,T) \cdot T \tau_{\mu\nu} \mathcal{P}^{\mu\nu}}
\,,\label{eq:eta_gen}
\end{equation}
with $ \mathcal{P}^{\mu\nu}$ denoting the second moment of the collision term 
\begin{equation}
\mathcal{P}^{\mu\nu} = \int p^\mu p^\nu \mathcal{C}[f] d\Gamma \,.\label{eq:pmunu}
\end{equation}
An obvious property of the obtained expression is the explicit dependence on the shear
tensor components. But since the shear viscosity coefficient is a medium
property and should not depend on the dynamics encoded in $\tau^{\mu\nu}$, Eq. (\ref{eq:eta_gen})
should be considered in the limit $\tau^{\mu\nu}\to 0$. 

The functional $\mathcal{P}^{\mu\nu}$ contains information about the collisional processes in the
medium. We consider only local interactions between constituents of the medium and assume them to
be Boltzmann-particles, i.e. quantum statistics effects are neglected. If both elastic and
inelastic interactions are considered, the second moment of the collision term takes the following
form:
\begin{eqnarray}
&&\mathcal{P}^{\mu\nu} =\int d\Gamma p^\mu p^\nu \mathcal{C} = \nonumber \\
&&\frac{1}{2} \int dw_1 dw_2 \frac{1}{2} \int dw^\prime_1
dw^\prime_2
p_1^\mu p_1^\nu f^\prime_1 f^\prime_2 \vert M_{1^\prime 2^\prime \to 12} \vert^2 (2\pi)^4
\delta^{(4)}
(p^\prime_1+p^\prime_2-p_1-p_2) - \nonumber\\
&&\frac{1}{2} \int dw_1 dw_2 \frac{1}{2} \int dw^\prime_1
dw^\prime_2
p_1^\mu p_1^\nu f_1 f_2 \vert M_{1 2 \to 1^\prime 2^\prime} \vert^2 (2\pi)^4
\delta^{(4)}
(p^\prime_1+p^\prime_2-p_1-p_2) + \nonumber\\
&&\frac{1}{2} \int dw_1 dw_2  dw_3 \frac{1}{2} \int dw^\prime_1
dw^\prime_2
p_1^\mu p_1^\nu f^\prime_1 f^\prime_2 \vert M_{1^\prime 2^\prime \to 123} \vert^2 (2\pi)^4
\delta^{(4)}
(p^\prime_1+p^\prime_2-p_1-p_2-p_3) + \nonumber\\
&+& \int dw_1 dw_2   \frac{1}{6} \int dw^\prime_1 dw^\prime_2
dw^\prime_3 p_1^\mu p_1^\nu f^\prime_1
f^\prime_2 f^\prime_3 \vert M_{1^\prime 2^\prime 3^\prime \to 12} \vert^2 (2\pi)^4 \delta^{(4)}
(p^\prime_1+p^\prime_2 + p^\prime_3-p_1-p_2)- \nonumber\\
&-& \frac{1}{2}\int dw_1 dw_2 dw_3  \frac{1}{2} \int dw^\prime_1
dw^\prime_2 
p_1^\mu p_1^\nu f_1 f_2 f_3 \vert M_{1 2 3 \to 1^\prime 2^\prime} \vert^2 (2\pi)^4 \delta^{(4)}
(p_1+p_2 + p_3-p^\prime_1-p^\prime_2) - \nonumber\\
&-& \int dw_1 dw_2   \frac{1}{6} \int dw^\prime_1 dw^\prime_2 dw^\prime_3
 p_1^\mu p_1^\nu f_1
f_2 \vert M_{12 \to 1^\prime 2^\prime 3^\prime} \vert^2 (2\pi)^4 \delta^{(4)}
(p_1+p_2-p^\prime_1-p^\prime_2 - p^\prime_3) \nonumber\\
\label{eq:collterm}
\end{eqnarray}
with $dw\equiv \frac{1}{2} d\Gamma = \frac{1}{2 (2\pi)^3} \frac{d^3p}{E}$.
The first two terms in (\ref{eq:collterm}) account for elastic $2\to 2$ processes, whereas the last
four terms capture the inelastic $2\to 3$ and reverse $3\to 2$ processes. The numerical factors
$1/2$ and $1/6$ account for multiple counting of identical Boltzmann particles. 

In order to evaluate the integrals in (\ref{eq:collterm}) analytically, the distribution functions
will be replaced by the near-equilibrium approximations (\ref{eq:grad}) and only terms up to first
order in $\phi$ will be considered:
\begin{equation}
 \prod_i f_i \approx \left( 1 + \sum_j\phi_j \right) \cdot \prod_i f_{0,i} \,.
\end{equation}
The equilibrium distribution $f_0$ is the Boltzmann distribution
\begin{equation}
 f_0 = d \cdot e^{ - p_\alpha u^\alpha / T + \mu/T }  \,,\label{eq:f0}
\end{equation}
with the degeneracy factor $d$. 

Analytic evaluation of the integrals in (\ref{eq:collterm}) can be simplified if a specific
symmetry for the geometry of the system is considered. Since the shear viscosity coefficient is a
material property, its value cannot depend on the form of the energy-momentum tensor distortion,
i.e. any plausible form for the shear tensor can be considered to evaluate (\ref{eq:collterm}). The
simplest geometry that can be considered for the shear tensor, which has to be traceless, is a
diagonal from with one independent component $\tau$:
\begin{equation}
 \tau_{\mu\nu} = diag(0,\tau/2,\tau/2,-\tau)\,.
\label{eq:1dshear}
\end{equation}
This form of the shear tensor implies deviation from isotropy in $z$-direction and isotropy in the
transverse plane. Systems with this kind of symmetry have been studied e.g. in Refs.
\cite{Bouras:2010hm,El:2010mt}. With the shear tensor given by Eq. (\ref{eq:1dshear}), the
expression for the shear viscosity (\ref{eq:eta_gen}) as well as
the collision term moment (\ref{eq:collterm}) are significantly simplified, since now $\tau$ is the
only unknown parameter, for which the $\tau\to 0$ limit must be considered. In fact, with the
introduced form of the shear pressure tensor the correction (\ref{eq:phi}) to the distribution
function becomes (for sake of compactness we will suppress the dependence of $c(\cdot)$ on $e,p,T$
in the notation)
\begin{equation}
\phi(x,p) = c \cdot \tau \cdot \left( \frac{1}{2} p_T^2 - p_z^2 \right) \,.
\label{eq:phi_1} 
\end{equation}
Now the contraction $\tau_{\mu\nu} \mathcal{P}^{\mu\nu}$ in (\ref{eq:eta_gen}) can be rewritten as
follows 
\footnotesize
\begin{eqnarray}
&&\tau_{\mu\nu}\mathcal{P}^{\mu\nu} =\tau_{\mu\nu} \int d\Gamma p^\mu p^\nu \mathcal{C} = \nonumber \\
&&\frac{\tau}{4} \int dw_1 dw_2  dw^\prime_1 dw^\prime_2
\left( \frac{1}{2} p_{1,T}^2  - p_{1,z}^2 \right) \cdot f^\prime_{0,1} f^\prime_{0,2} \times
\nonumber\\
&&\times \left(1 + c\cdot\tau\cdot (\frac{1}{2} p^{\prime 2}_{1,T} -
p^{\prime 2}_{1,z}) + c\cdot\tau\cdot (\frac{1}{2} p^{\prime 2}_{2,T} -
p^{\prime 2}_{2,z}) \right) \cdot  \vert M_{1^\prime 2^\prime \to 12} \vert^2 (2\pi)^4
\delta^{(4)} (..) - \nonumber\\
&&\frac{\tau}{4} \int dw_1 dw_2 dw^\prime_1 dw^\prime_2 
\left( \frac{1}{2} p_{1,T}^2  - p_{1,z}^2 \right) \cdot f_{0,1} f_{0,2} \times \nonumber\\ 
&& \times \left(1 + c\cdot\tau\cdot (\frac{1}{2} p_{1,T}^2 - p_{1,z}^2) + c\cdot\tau\cdot (\frac{1}{2}
p_{2,T}^2 - p_{2,z}^2) \right) \cdot \vert M_{1 2 \to 1^\prime 2^\prime} \vert^2 (2\pi)^4
\delta^{(4)} (..) + \nonumber\\
&&\frac{\tau}{4} \int dw_1 dw_2  dw_3 dw^\prime_1 dw^\prime_2
\left( \frac{1}{2} p_{1,T}^2  - p_{1,z}^2 \right) f^\prime_{0,1} f^\prime_{0,2} \times \nonumber\\
&&\times \left( 1 +  c\cdot\tau\cdot (\frac{1}{2} p^{\prime 2}_{1,T} -
p^{\prime 2}_{1,z}) + c\cdot\tau\cdot (\frac{1}{2} p^{\prime 2}_{2,T} - p^{\prime 2}_{2,z})
\right) \cdot \vert M_{1^\prime 2^\prime \to 123} \vert^2 (2\pi)^4
\delta^{(4)} (..) + \nonumber\\
&& \frac{\tau}{6}\int dw_1 dw_2 dw^\prime_1 dw^\prime_2 dw^\prime_3 
\left( \frac{1}{2} p_{1,T}^2  - p_{1,z}^2 \right) \cdot 
f^\prime_{0,1} f^\prime_{0,2} f^\prime_{0,3} \times\nonumber\\
&&\times \left( 1 +  c\cdot\tau\cdot (\frac{1}{2} p^{\prime 2}_{1,T} -
p^{\prime 2}_{1,z}) + c\cdot\tau\cdot (\frac{1}{2} p^{\prime 2}_{2,T} - p^{\prime 2}_{2,z}) +
c\cdot\tau\cdot (\frac{1}{2} p^{\prime 2}_{3,T} - p^{\prime 2}_{3,z}) \right) \cdot \vert
M_{1^\prime 2^\prime 3^\prime \to 12} \vert^2 (2\pi)^4 \delta^{(4)} (..)- \nonumber\\
&& \frac{\tau}{4}\int dw_1 dw_2 dw_3 dw^\prime_1 dw^\prime_2 
\left( \frac{1}{2} p_{1,T}^2  - p_{1,z}^2 \right) \cdot f_{0,1} f_{0,2} f_{0,3}
\times\nonumber\\
&&\times \left( 1 + c\cdot\tau\cdot (\frac{1}{2} p^2_{1,T} -
p^2_{1,z}) + c\cdot\tau\cdot (\frac{1}{2} p^2_{2,T} - p^2_{2,z}) +
c\cdot\tau\cdot (\frac{1}{2} p^2_{3,T} - p^2_{3,z}) \right) \cdot 
\vert M_{1 2 3 \to 1^\prime 2^\prime} \vert^2 (2\pi)^4 \delta^{(4)} (..) - \nonumber\\
&& \frac{\tau}{6} \int dw_1 dw_2 dw^\prime_1 dw^\prime_2 dw^\prime_3
\left( \frac{1}{2} p_{1,T}^2  - p_{1,z}^2 \right) \cdot f_{0,1} f_{0,2} \times \nonumber \\ 
&& \left(1 + c\cdot\tau\cdot (\frac{1}{2} p_{1,T}^2 - p_{1,z}^2) + c\cdot\tau\cdot (\frac{1}{2}
p_{2,T}^2 - p_{2,z}^2) \right) \cdot \vert M_{12 \to 1^\prime 2^\prime 3^\prime} \vert^2 (2\pi)^4
\delta^{(4)} (..) \nonumber\\
\label{eq:momentofcollterm}
\end{eqnarray}
\normalsize
The parts of the integrals that do not contain $c$ (i.e. those parts where the
integration goes over $f_0$ and not $f_0\cdot \phi$) have to vanish, since there is no entropy
production in an equilibrated medium. The remaining integrals will then be of order $\tau^2$, just as
the expression in the numerator of (\ref{eq:eta_gen}), so that the shear viscosity coefficient will
not depend on $\tau$, as it should be.

\section{Evaluation of the kinematic integrals}\label{sec:evaluation}

In general the matrix elements $\vert M_{n\to m} \vert $ will have a highly non-trivial dependence
on momenta of the colliding particles and analytic evaluation might be difficult. Here we will
consider angle-independent matrix elements to obtain the shear viscosity as function
of total cross section. The total cross sections $\sigma_{22}$ and $\sigma_{23}$ for elastic resp.
inelastic two-particle processes can be defined according to \cite{byckling73,Xu:2004mz}:
\begin{align}
\sigma_{22} &= \frac{1}{2s} \frac{1}{2} \int  dw_1 \int dw_2 \vert M_{1^\prime 2^\prime \to 1 2}
\vert^2 (2\pi)^4 \delta\left( p_1 + p_2 - p^\prime_1 - p^\prime_2 \right)  \,,\label{eq:cs22}\\
\sigma_{23} &= \frac{1}{2s} \frac{1}{6} \int  dw_1 \int dw_2 \int dw_3 \vert M_{1^\prime 2^\prime
\to 1 2 3} \vert^2 (2\pi)^4 \delta\left( p_1 + p_2  + p_3 - p^\prime_1 - p^\prime_2 \right) 
\,.\label{eq:cs23}
\end{align}
For these definitions the center-of-mass energy squared of the colliding particles $s = (p_1 +
p_2)^2$ has been introduced. For a process with $3$ particles in the initial state a classical cross
section cannot be defined, but one can define an analogous quantity \cite{Xu:2004mz}
\begin{equation}
 \mathcal{I}_{32} = \frac{1}{2} \int  dw_1 \int dw_2 \vert M_{1^\prime
2^\prime 3^\prime \to 1 2} \vert^2 (2\pi)^4 \delta\left( p_1 + p_2 - p^\prime_1 - p^\prime_2
 - p^\prime_3\right) \,.\label{eq:I32}
\end{equation}
The squared matrix elements $\vert M_{2\to 3} \vert^2$ and $\vert M_{3\to 2} \vert^2$ depend on each
other by virtue of the detailed balance requirement:
\begin{equation}
 \vert M_{2\to 3} \vert^2 = d \cdot \vert M_{3\to 2} \vert^2 \,.
\end{equation}

If the matrix elements are angle-independent, i.e. if the scattering is \textit{isotropic}, as
considered for the following derivations, one can obtain simple relations between them, the cross
sections $\sigma_{22}$,  $\sigma_{23}$ and the quantity $\mathcal{I}_{32}$ \cite{Xu:2004mz} by
performing
the momentum integration in the definitions (\ref{eq:cs22}) -- (\ref{eq:I32}). Note that the
right-hand-sides of Eqs. (\ref{eq:cs22}) -- (\ref{eq:cs23}) are Lorentz-invariant and thus can be
conveniently evaluated in the center-of-mass frame of colliding particle pair:
\begin{align}
\vert M_{2\to 2} \vert^2 &= 32 \pi \cdot s \cdot \sigma_{22} \,,\label{eq:M22}\\
\vert M_{2\to 3} \vert^2 &= 192 \pi^3 \cdot \sigma_{23} \,,\label{eq:M23}\\
\mathcal{I}_{32} &= \frac{192}{d} \pi^2 \sigma_{23} \,.\label{eq:I}
\end{align}
Here, again, $d$ is the degeneracy factor of the constituent particles.  

With the definitions (\ref{eq:cs22}) -- (\ref{eq:I32}) half of the terms (the second, fifth and
sixth terms) in (\ref{eq:collterm}) are significantly simplified. Calculation of the remaining
terms is a bit more involved. The first, third and fourth terms in (\ref{eq:collterm}) contain
integrals over the $n=2$
or $3$ \textit{final state momenta} $dw_1 \ldots dw_n$ which can be formally written as
\begin{equation}
 \int dw_1 \ldots\int dw_n \mathcal{G}(p_1\ldots p_n) \vert M_{m\to n}\vert^2 \delta ( \sum_m p_{\rm
initial} - \sum_n p_{\rm final} ) \,.\label{eq:int}
\end{equation}
In the latter expression $\mathcal{G}$ denotes a (polynomial) function of the final state momenta.
For the matrix elements the relations  (\ref{eq:M22}) -- (\ref{eq:I}) can be used. However it is
important to note that the expression (\ref{eq:int}) is not anymore Lorentz-invariant
due to presence of the function $\mathcal{G}$. At least for the terms describing a two-particle
initial state it is still convenient to transform the integral into the center-of-mass frame for
evaluation and to apply a back-transform afterwards. This means rewriting Eq. (\ref{eq:int}) in
the following form
\begin{equation}
 \int dw^*_1 \ldots\int dw^*_n \mathcal{G}(p^*_{TR,1}\ldots p^*_{TR,n}) \vert M_{m\to n}\vert^2
\delta ( \sum_m p^*_{\rm initial} - \sum_n p^*_{\rm final} ) \,.\label{eq:int_TR}
\end{equation}
where $*$ denotes quantities in  the center-of-mass frame, in which $E^*_1+E^*_2 = \sqrt{s}$ and
$\vec p_1^*+\vec p_2^*=0$. To
rewrite (\ref{eq:int}) into the form (\ref{eq:int_TR}) Lorentz invariance of $dw_i$ and of the
4-dimensional Delta-function can be used. Notation $\mathcal{G}(p^*_{TR,1}\ldots p^*_{TR,n})$
means that function $\mathcal{G}(\cdotp)$ acts on momenta that are Lorentz-transformed from the
center-of-mass frame into the lab frame:
\begin{align}
 E^*_{TR} &= \gamma(E^* + \vec{p^*} \vec \beta_{CM}) \,\\
\vec {p^*}_{TR} &= \vec{p^*} + \frac{\vec\beta_{CM}
\gamma_{CM}^2}{1+\gamma_{CM}}\vec\beta_{CM}\cdot \vec{p^*} + \gamma_{CM}\vec\beta_{CM} E^* \,.
\end{align}
with the Lorentz boosts
\begin{align}
\vec\beta_{CM} &= \frac{\vec p_1 + \vec p_2}{E_1+E_2} \,,\\
\gamma_{CM} &= \frac{1}{\sqrt{1-\vec\beta_{CM}^2}} \,.
\end{align}
For the term describing a three-particle initial state (the fourth term in (\ref{eq:collterm})
resp. (\ref{eq:momentofcollterm})) such transformation into and then back from the center-of-mass
frame does not induce any significant simplification of the calculation so that the integration must
be evaluated directly in the lab frame.

After evaluating all the integrals we finally obtain the following expression for the shear
viscosity coefficient:
\begin{equation}
\eta = \frac{6}{5}\cdot\frac{T}{\sigma_{22} + \frac{6}{5}\sigma_{23} + 
\frac{3}{10}\lambda\sigma_{23}} \,.\label{eq:viscoGEN}
\end{equation}
The second and the third terms in the denominator of the latter equation account for contributions
of $2\to 3$ resp. $3\to 2$ processes. The factor $\lambda$ in the $3\to 2$ term denotes the
fugacity, which is a measure for the degree of saturation of the particle density in a system 
\begin{equation}
\lambda = n/n_{eq} \,,
\end{equation}
with $n_{eq}=d\cdot \int e^{-u_\mu p^\mu/T} d^3 p$. Setting $\lambda=1$ (or, accordingly, $\mu=0$
in Eq. (\ref{eq:f0})), i.e. considering a chemically equilibrated system, we obtain
\begin{equation}
\boxed{
\eta = \frac{6}{5}\cdot\frac{T}{\sigma_{22} + \frac{3}{2}\sigma_{23}} \label{eq:visco}
}
\end{equation}
Neglecting the contribution from inelastic scatterings, this formula is identical with the
already known results from Refs. \cite{DeGroot} and \cite{Denicol:2012cn}.

\section{Comparison with kinetic transport calculations}\label{sec:comparison}

In order to confirm the validity of Eq.~\eqref{eq:visco} it is necessary to extract
the transport coefficients from the full solution of the relativistic Boltzmann equation.
An accurate numerical solver of the relativistic Boltzmann equation
is the Boltzmann Approach to Multi-Parton Scatterings
(BAMPS) \cite{Xu:2007aa,Xu:2004mz}. This has been demonstrated in simulating
shock waves in various scenarios \cite{Bouras:2009nn,Bouras:2012mh}. The success of those
approaches then motivated to compare different approaches of dissipative hydrodynamics \cite{Bouras:2010hm,Denicol:2012vq}
and relativistic lattice Boltzmann calculations \cite{oai:arXiv.org:1009.0129,oai:arXiv.org:1109.0640} to the numerical
solutions of BAMPS in order to verify their validity.

The accurate solution of the relativistic Boltzmann equation allows the numerical
extraction of the transport coefficient like the shear viscosity and heat conductivity
from BAMPS. This has been demonstrated using the Green-Kubo formalism, which has
been successfully applied to extract the shear viscosity to a very high precision
from BAMPS calculations \cite{Wesp:2011yy} and agreed excellently with extraction
of the shear viscosity coefficient using the classical picture of a velocity gradient \cite{Reining:2011xn}.
Furthermore, in a recent work the extraction of heat conductivity from BAMPS calculations
could clarify which theoretical prediction of the heat conductivity coefficient each
originating from different derivations of dissipative hydrodynamics is the most accurate \cite{Greif:2013bb}.
In the following we apply the Green-Kubo formalism to extract the shear viscosity
from BAMPS calculations in order to verify the accuracy of Eq.~\eqref{eq:visco}.

The Green-Kubo relation allows to connect a linear transport coefficient
with the correlation function of its current.
The evaluation of the shear viscosity using the Green-Kubo method is realized
by calculating the correlator of the shear-stress tensor in a thermal and equilibrated
without any spatial gradients.

The extraction of the shear-stress tensor from
BAMPS is evaluated by calculating the energy-momentum tensor defined as
\begin{equation}
T^{\mu \nu }(t) = \int d\Gamma \, p^{\mu}p^{\nu } f(p,t).
\label{eq:tmunu_general_BAMPS}
\end{equation}
In BAMPS this is realized by summing up over all discrete particle momenta
in the volume averaged simulation
\begin{equation}
T^{\mu \nu }(t) =\frac{1}{V N_{\rm test}}\sum\limits_{i=1}^{N}\frac{p_{i}^{\mu
}p_{i}^{\nu }}{p_{i}^{0}},  \label{eq:tmunu_BAMPS}
\end{equation}
Here, $N$ is the number of particles, while $V$ is the volume of the system and
$N_{\rm test}$ is the test particle number \cite{Xu:2004mz,Bouras:2010hm}. Due to the
fact that we consider a stationary system, the corresponding fluctuations of
the energy-momentum tensor are of the same order as the one of the shear-stress
tensor and therefore the shear-stress tensor has not to be calculated explicitly.
This is realized in the fact that the correlation function of energy-momentum tensor
is the same as the one from the shear-stress tensor in equilibrium:
\begin{equation}
	\langle T^{xy}(t) T^{xy}(0) \rangle \ =  \langle \tau^{xy}(t) \tau^{xy}(0) \rangle \,,
\end{equation}
with the correlation function:
\begin{equation}
	 \langle \tau^{xy}(t) \tau^{xy}(0) \rangle = \int ds \, \tau^{xy}(t+s) \tau^{xy}(s)
\end{equation}
Finally, with a thermal prefactor, the corresponding Green-Kubo relation for the shear
viscosity has the following form \cite{Wesp:2011yy}:
\begin{equation} \label{green_kubo_definition} 
	\eta = \frac{V}{10 \text{T}} \int_{0}^{+ \infty} \mathrm{d}t \ \langle \tau^{ij}(t) \tau^{ij}(0)
	\rangle \,,
\end{equation}
where $\langle \cdots \rangle$ denotes the ensemble averaged correlation function in thermal equilibrium
of the shear components at time $t=0$ and at $t$. The correlator is summed over all spatial components
$i$ and $j$. In case of stochastic processes, the correlation function in Eq.~\eqref{green_kubo_definition} will have the form of an exponential.
More details to the Green-Cubo method and the extraction within BAMPS is shown explicitly in \cite{Wesp:2011yy}.

\begin{figure}
\includegraphics[width=11.5cm]{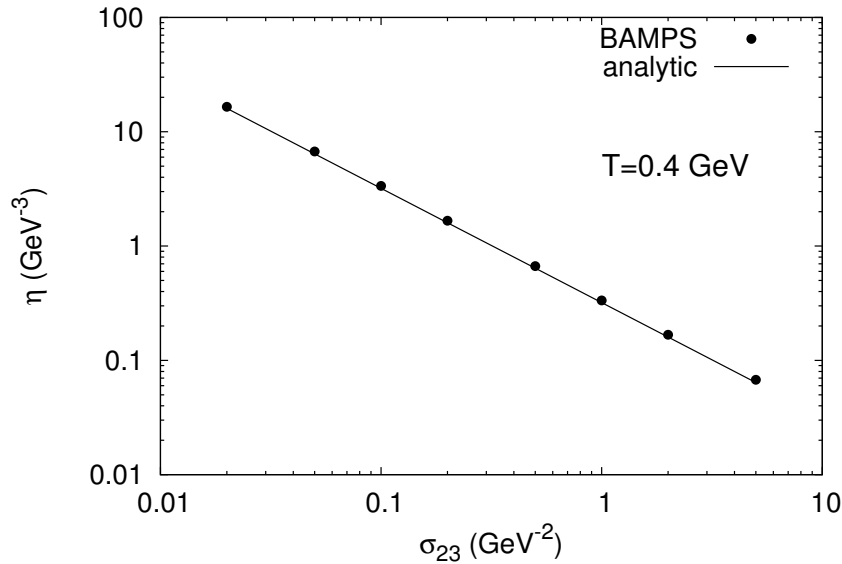}
\caption{Shear viscosity in presence of inelastic processes only, extracted from BAMPS using
Green-Kubo formalism (symbols) and calculated using Eq. (\ref{eq:visco}). The
results are obtained employing angle-independent (isotropic) differential cross sections.}
\label{fig:eta}      
\end{figure}

In Fig. \ref{fig:eta} we demonstrate the shear viscosity values
extracted this way from BAMPS calculations for which only inelastic $2\to 3$ and $3\to 2$ processes
were considered. The shear viscosity is shown as function of the inelastic cross section
$\sigma_{23}$. We observe a very good agreement of the results extracted from BAMPS (symbols) with
the results obtained using Eq. (\ref{eq:visco}) (lines)
derived in this work. We also would like to mention that Eq. (\ref{eq:visco}) confirms the
values of the shear viscosity which we obtained in Ref. \cite{El:2010mt} by numerical evaluation of
the Equation (\ref{eq:eta_gen}). Equation (\ref{eq:visco}) implies that if the total cross
sections are equal, relaxation of the medium towards equilibrium will proceed $1.5$ times
faster if only inelastic processes are involved as compared to the case when only elastic processes
are involved (since the hydrodynamic relaxation time is directly proportional to the shear
viscosity coefficient -- comp. \cite{Muronga:2003ta,El:2009vj}). This reflects the fact that the
inelastic processes are more efficient in driving the medium towards equilibrium, as was
demonstrated in Refs. \cite{Xu:2007aa,El:2008yy,El:2010mt}.

\section{Conclusions}\label{sec:conclusion}

We have derived an analytic expression relating the shear viscosity coefficient of an
ultra-relativistic gas to the total elastic and inelastic cross sections of isotropic scattering
processes in the medium. From the obtained expression one recognizes that, given all total cross
sections are equal, isotropic inelastic processes are $1.5$ times more efficient than the elastic
ones in restoring kinetic equilibrium in a system. The values of the shear viscosity
obtained by Eq. (\ref{eq:visco}) are in very good agreement with the results of
Green-Kubo based calculations performed using the kinetic transport model BAMPS. The correspondence
between shear viscosity and total cross sections of elastic and inelastic processes reported is
another step to establish a connection between the kinetic transport and dissipative hydrodynamic
models.

\section*{Acknowledgements}

This work was supported by the Helmholtz International Center for FAIR within the framework of the
LOEWE program launched by the State of Hesse. BAMPS calculations were performed using the
resources of the Center for Scientific Computing of the Goethe University Frankfurt.

\end{document}